# Static and dynamic magnetic properties of K$_3$CrO$_4$

Liliia D. Kulish,* Graeme R. Blake

*l.kulish@rug.nl

*Zernike Institute for Advanced Materials, University of Groningen, Nijenborgh 4, 9747 AG, Groningen, the Netherlands*

**ABSTRACT:** We report on the magnetic properties of geometrically frustrated K$_3$CrO$_4$, in which Cr$^{5+}$ cations are arranged on a distorted pyrochlore lattice. The crystal structure, static and dynamic magnetic properties of the compound are investigated in detail. A combination of DC and AC magnetic susceptibility measurements together with thermoremanent magnetization decay measurements reveal several magnetic transitions: the onset of glassy canted antiferromagnetic order occurs at 36 K, followed by the appearance of ferromagnetic/ferrimagnetic cluster glass behavior below the freezing temperature of 20 K. Further field-induced, temperature-dependent transitions are observed in the range 3-10 K. The frequency dependence of the freezing temperature for the cluster glass state is analyzed on the basis of dynamic scaling laws including the critical slowing down formula and the Vogel-Fulcher law.

## 1. INTRODUCTION

Geometrically induced magnetic frustration can promote the existence of various exotic magnetic states in a single compound. A manifold of ground states that are similar in energy is often manifested by spin ice / liquid behavior as well as by the emergence of exotic magnetic structures such as helical / cycloidal spirals and periodic states with non-trivial topologies composed of skyrmions and antiskyrmions [1-3]. Magnetic skyrmions are of particular interest because they can be controlled by extremely small electric currents, hence there are promising perspectives for application in novel spintronic and information storage devices [4,5]. The material science of skyrmions in frustrated magnets is expected to be highly rich: skyrmions and antiskyrmions can co-exist in one material, which gives a new dimension to logic operations that can be performed with these objects. Furthermore, in frustrated magnets, the skyrmion helicity becomes a new collective degree of freedom coupled to the skyrmion motion. Additionally, frustrated magnets are expected to exhibit smaller skyrmions (a few nm in size) compared to chiral magnets which comprise most of the known skyrmion host materials; this has the potential to lead to higher information density [1,3,6].

A main challenge in this field is thus to find frustrated magnets that host skyrmions. A promising compound from this point of view is K$_3$CrO$_4$. It is a 3d$^1$ system (S = ½) with the unusual 5+ oxidation state of chromium. K$_3$CrO$_4$ has a cubic structure with the same chiral space group $P2_13$ as the well-studied skyrmionic materials MnSi [7] and Cu$_2$OSeO$_3$ [8].

In the literature, different synthesis routes have been described [9,10], including a recent investigation in which *in situ* high-temperature X-ray diffraction measurements were performed during the reduction of the Cr$^{6+}$ compound K$_2$CrO$_4$, which led to the formation of different K$_x$CrO$_y$ compounds including K$_3$CrO$_4$ [11].

In addition, it has been shown that single crystals of a β-K$_3$CrO$_4$ polymorph can be grown from the original cubic polymorph when heated at 180 ˚C for 1 week. β-K$_3$CrO$_4$ adopts a tetragonal structure with space group $I\bar{4}2m$ and shows paramagnetic behavior to below 5 K [9]. The magnetic behavior of cubic K$_3$CrO$_4$ has not been reported.

Examples of compounds containing Cr$^{5+}$ ions are scarce in the literature. There are perchromate compounds M$_3$CrO$_8$ (M is an alkali metal cation) with ferroelectric properties, where Cr is coordinated in eight-fold fashion by peroxide (O$_2^{2-}$) ions in a dodecahedral configuration [12]. A combined EPR and magnetic susceptibility analysis demonstrated the dynamics of the electron spin exchange and antiferromagnetic exchange coupling in K$_3$CrO$_8$ single crystals [13]. The influence of the Cr$^{5+}$ ions on the magnetic properties of YbCrO$_4$ was investigated by both bulk magnetic measurements and $^{170}$Yb Mössbauer spectroscopy. Antiferromagnetic coupling between the Yb$^{3+}$ and Cr$^{5+}$ sublattices leads to ferrimagnetic ordering below 25 K, driven by exchange within the chromium sublattice [14].

Here, we explore the magnetic behavior of K$_3$CrO$_4$ for the first time. We utilize a simple and fast reduction reaction to obtain the pure sample. Powder X-ray diffraction shows that the cubic polymorph is obtained with space group $P2_13$. We use DC and AC magnetic susceptibility studies together with thermoremanent magnetization decay measurements to reveal that K$_3$CrO$_4$ undergoes a series of magnetic phase transitions with temperature. A paramagnetic to glassy canted antiferromagnetic transition takes place at ~ 36 K, and the sample becomes glassy in nature below a freezing temperature of 20 K. The frequency dispersion of the temperature-dependent AC susceptibility is described by dynamic scaling theory and the Vogel-Fulcher law, which identifies it as a ferromagnetic/ferrimagnetic cluster glass state. In addition, further field-induced temperature-dependent transitions are detected at lower temperatures of 3-10 K.

## 2. MATERIALS AND METHODS

A polycrystalline $K_3CrO_4$ sample was prepared by the reduction of potassium chromate ($K_2CrO_4$) by heating in flowing hydrogen, based on the process earlier described by Liang *et al.* [15]. Pre-ground $K_2CrO_4$ was placed in an alumina boat crucible with an alumina cap, which was inserted in a tube furnace. A purified gas mixture ($H_2$ + Ar, 15% + 85%) was then introduced at a constant flow rate of 200 ml/min into the tube. The temperature was raised to 400 °C at 10 °C/min, and then to 450 °C at 1 °C/min. The sample was held at 450 °C for 30 min, and then the furnace was naturally cooled to ambient temperature. The sample was immediately transferred to a nitrogen-filled glove-box and kept in an inert atmosphere with $O_2$ and $H_2O$ concentrations of less than 10 ppm due to the extreme air-sensitivity of the obtained compound.

The reduction process that occurs on heating was confirmed by means of simultaneous thermogravimetric analysis (TG) and differential scanning calorimetry (DSC) on a TG 2960 SDT instrument using a hydrogen-argon flow ($H_2$ / Ar, 15% / 85%) of 100 mL/min; the heating rate was 5 °C/min over the temperature range 30 °C to 1000 °C (Fig. S1). The obtained data are in good agreement with a previous report on the reduction of $K_2CrO_4$ [16].

The phase purity and crystal structure of the product were determined by X-ray powder diffraction (XRD) using a Bruker D8 Advance diffractometer operating with Cu Kα radiation in the 2θ range 10-70 °. A rotating glass capillary (diameter 0.5 mm) containing the sealed sample was used in transmission geometry. The XRD data were fitted by Rietveld refinement using the GSAS software [17]. Magnetic measurements were performed on a Quantum Design MPMS SQUID-based magnetometer. Magnetic susceptibility scans were performed on warming over the range 2-400 K, and magnetization versus applied field curves were obtained between -7 T and 7 T at 3-60 K. AC susceptibility measurements were performed using a 3.8 Oe oscillating field superimposed on different DC fields: 0, 200 and 400 Oe. Thermoremanent magnetization decay experiments were performed by applying a 1 T field, cooling the sample to 10 K at 10 K/min, then cooling to 3 or 7 K at 2 K/min (below the glass freezing temperature). After 30 sec, the field was removed and the magnetization was measured as a function of time.

## 3. RESULTS AND DISCUSSION

### 3.1. Structural characterization

Structural analysis of the $K_3CrO_4$ sample using powder XRD shows a single-phase product for which the peak positions are consistent with the previously reported cubic space group $P2_13$ [9]. The refined lattice parameter is $a = 8.3158(10)$ Å and the unit cell volume is $V = 575.05(22)$ Å$^3$ (the fitted XRD data are shown in Fig. S2, the atomic coordinates are listed in Table S1). In this structure chromium atoms occupy a single crystallographic position and are tetrahedrally coordinated by oxygen (Fig. 1(a)). The $CrO_4$ tetrahedra are not directly connected to each other, thus Cr-Cr magnetic exchange interactions take place via two O atoms. In addition, it should be mentioned that the chromium atoms when envisaged alone form a slightly distorted pyrochlore structure, which is geometrically frustrated and thus has possible consequences for the magnetic properties. The general chemical formula of a pyrochlore is $A_2B_2X_7$, where the A and B cations form corner-sharing, interpenetrating [$A_4$] and [$B_4$] tetrahedra. In the case of $K_3CrO_4$, Cr occupies both the A and B sites giving rise to the configuration shown in Fig. 1(b).

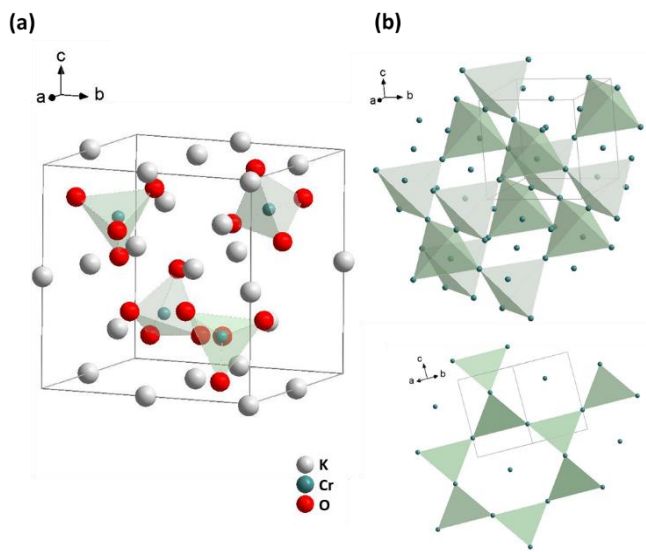

Fig. 1. (a) Crystal structure of $K_3CrO_4$. The potassium, chromium, and oxygen atoms are represented by white, dark green, and red spheres respectively. The unit cell is outlined. (b) Schematic representation of the Cr-sublattice, which forms a slightly distorted pyrochlore structure.

## 3.2. DC magnetic susceptibility

The magnetic properties of K$_3$CrO$_4$ were first investigated by performing DC magnetization versus temperature measurements. Zero-field-cooled (ZFC) and field-cooled (FC) measurements were performed in applied magnetic fields of 200 Oe and 10 kOe on warming over the temperature range 2-400 K. A bifurcation of the FC and ZFC curves below a characteristic temperature ($T_{irr}$ = 38 K), with a well-defined peak in the ZFC branch ($T_g$ = 30 K), is observed under 200 Oe DC field (Fig. 2(a)). Such splitting shows a dependence on the thermal-magnetic history of the sample and can arise from a variety of phenomena such as glassy, spin ice / liquid, superparamagnetic, disordered antiferromagnetic, and spin-spiral states [18,19]. The FC magnetization shows a continuous increase with decreasing temperature. At the same time the ZFC magnetization crosses zero and becomes negative below 14 K. Negative magnetization can appear in complex ferrimagnetic or canted antiferromagnetic systems [20]. There is also a maximum at 3 K in both the ZFC and FC curves (Fig. 2(a), inset; Fig. S3). However, measurement in a higher applied DC field of 10 kOe leads to different magnetic behavior of the sample below 60 K. Both ZFC and FC curves exhibit a continuous increase with decreasing temperature and there is only a tiny degree of ZFC-FC splitting below 8 K. No sign of the magnetic transition at 30 K is observed; there are only two broad maxima at 3 K and 38 K (Fig. S4).

The inverse susceptibility of the sample (Fig. 2(b)) has a linear part only in 10 kOe of applied DC field and only in FC mode (Fig. S5). These data were fitted according to the Curie-Weiss law, $\chi_{mol} = C/(T-\theta) + \chi_0$, where $C$ is the Curie constant and $\theta$ is the Weiss constant. The temperature-independent term $\chi_0$ consists of the sum of the diamagnetic contributions of the core electrons $\chi_{dia}$ (Cr$^{5+}$+3K$^+$+4O$^{2-}$) [21] and the van Vleck paramagnetic contribution of the Cr$^{5+}$ ion [22,23]. Fitting was performed above 250 K; the deviation from linearity at lower temperature is most likely due to short-range interactions. The extracted negative Weiss constant, $\theta$ = -172 K, implies predominant antiferromagnetic interactions. From the extracted Curie constant of 0.796 emu·K·mol$^{-1}$, an effective moment $\mu_{eff}$ of 2.52 $\mu_B$ per Cr atom is determined using the formula $C = \mu_{eff}^2/8$. The theoretical spin-only $\mu_{eff}$ of Cr$^{5+}$ is expected to be 1.73 $\mu_B$ ($\mu_{eff}$(Cr$^{3+}$) = 3.87 $\mu_B$; $\mu_{eff}$(Cr$^{2+}$) = 4.90 $\mu_B$) [18].

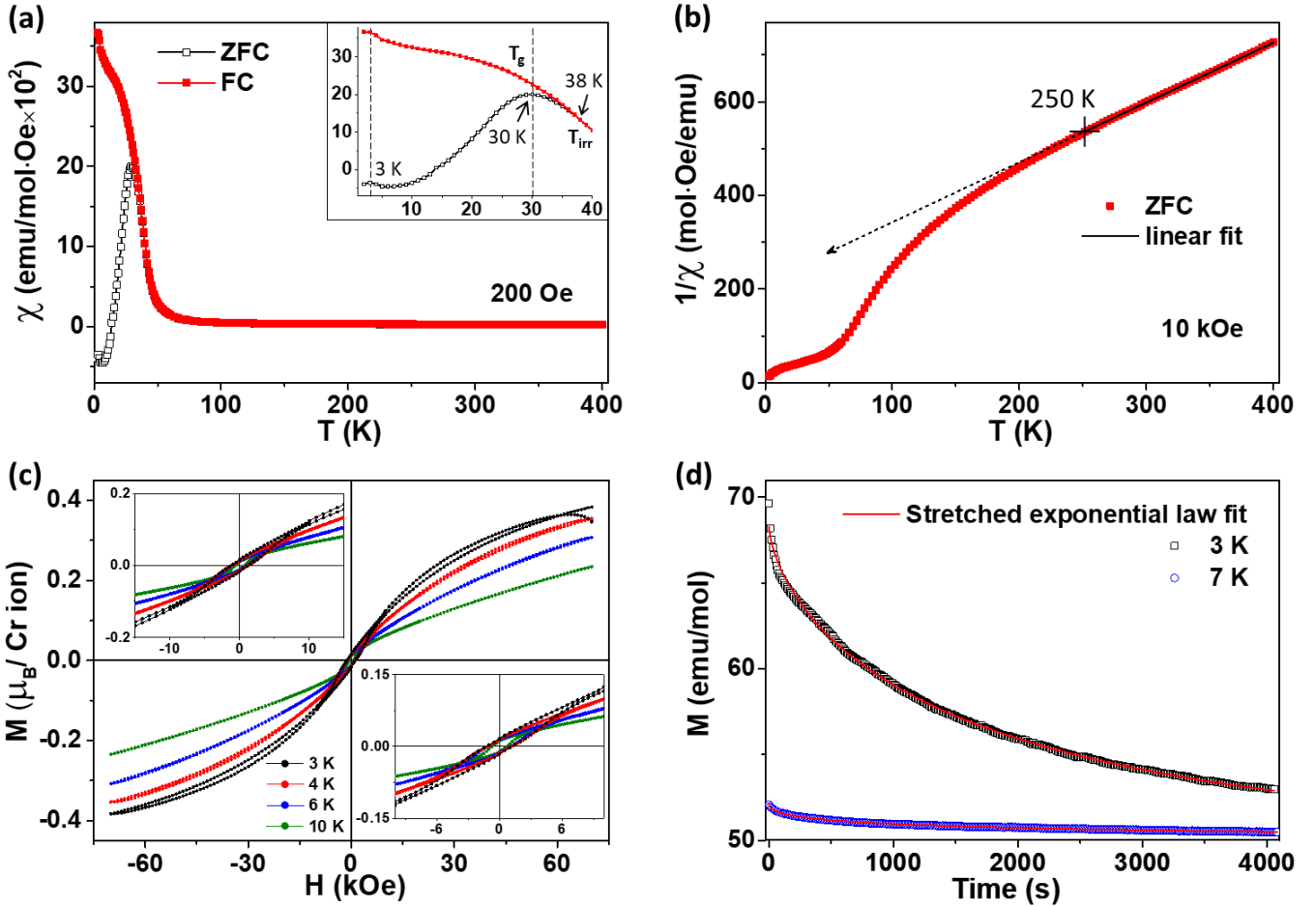

Fig. 2. (a) Temperature dependence of ZFC (open symbols) and FC (solid symbols) DC magnetic susceptibility of K$_3$CrO$_4$ measured on warming in a field of 200 Oe. The inset shows the DC susceptibility of the samples in the 2-40 K range. (b) FC inverse DC susceptibility of K$_3$CrO$_4$ as a function of temperature measured under 10 kOe. The line is a linear fit to the experimental data above 250 K using the Curie-Weiss law. (c) Magnetization vs applied DC field curves at 3-10 K for K$_3$CrO$_4$. The insets show closer views of the low-field region. (d) Magnetization decay as a function of time measured after cooling the sample under a 1 T field to 3 or 7 K and then removing the field. The curves are fits to the experimental data using the stretched exponential function (Eq. (3)).

The possible presence of magnetic frustration can be inferred from the frustration parameter $f = |\theta_{CW}|/T_C$ [24]. Although this parameter is strictly speaking only valid for a long-range-ordered state below a critical temperature $T_C$, which does not seem to be achieved in the case of $K_3CrO_4$ (see discussion of AC susceptibility data below), it can nevertheless provide a useful indicator of frustration if $T_C$ is replaced by $T_g$. For $K_3CrO_4$ $|\theta_{CW}|$ is 4.8 times greater than $T_g$ (36 K) implying a moderate level of frustration.

### 3.3. AC magnetic susceptibility

To further investigate the origin of the features observed in the ZFC curves (Fig. 2(a)), the temperature dependence of the AC susceptibility $\chi_{AC}$ was measured over the temperature range 8-50 K at seven different frequencies: 10, 50, 100, 250, 500, 750, 1000 Hz and in different applied DC fields: zero, 200 or 400 Oe (Fig. 3). AC measurements in the lower temperature range of 2.5-8 K (not shown) do not show any peaks in either the real or imaginary parts.

In the case of zero applied DC field (Fig. 3(a)), both the real component $\chi'(T)$ (reversible magnetization processes) and the imaginary part $\chi''(T)$ (losses due to irreversible processes) [25] exhibit frequency-dependent relaxation (marked as **1** in Fig. 3(a)). There is a single peak in $\chi'(T)$ which both decreases in height and shifts to higher temperature with increasing frequency. The corresponding peak in $\chi''(T)$ increases in height and shifts to higher temperature with increasing frequency. The temperature of this maximum (36 K at 10 Hz for $\chi'(T)$), which we refer to as $T_f$, is ~6 K higher than $T_g$ at which the peak in the ZFC DC susceptibility is observed. Moreover, $T_f$ approximately corresponds to the temperature at which the inflection point is observed in the imaginary part $\chi''(T)$. The peaks in $\chi'(T)$ and $\chi''(T)$ are both asymmetric, with an inflection point in $\chi'(T)$ at ~24 K.

The superposition of a DC field during the AC measurement leads to the appearance of another peak at a lower temperature for both the real and the imaginary parts (labelled as **2** in Fig. 3(b, c)), which might have developed from the shoulder in $\chi'(T)$ in Fig. 3(a). This transition at ~ 20 K is dynamic whereas peak **1** at 38 K now has no frequency dependence.

Peak **2** reflects field-induced magnetic relaxation, which is greatly enhanced with increasing applied DC field (Fig. 3). In addition, peak **2** shifts to lower temperature while peak **1** shifts to higher temperature with increasing DC field. Furthermore, the AC susceptibility of $\chi'(T)$ and $\chi''(T)$ for both transitions decreases rapidly with increasing DC field (Fig. S6).

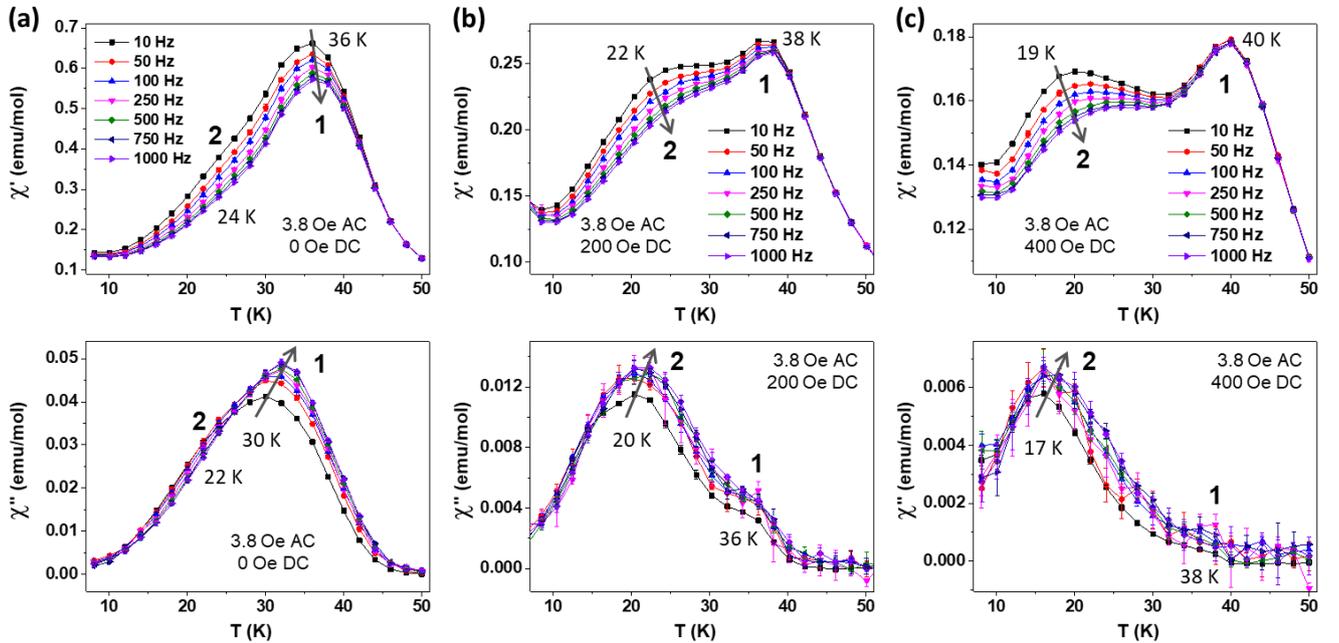

Fig. 3. Temperature dependence of the real and imaginary parts of the AC susceptibility in the temperature range 8-50 K at 10-1000 Hz frequency, measured using a 3.8 Oe oscillating field and different applied DC fields: (a) without DC field; (b) 200 Oe DC field; (c) 400 Oe DC field.

As mentioned above, the $Cr^{5+}$ ions are arranged on a distorted pyrochlore lattice. Therefore, it is possible that spin ice behavior typical for pyrochlores is exhibited by $K_3CrO_4$. Spin ices demonstrate a slowing down of the spin dynamics on cooling in similar fashion to spin-glass compounds, with a low-temperature peak in the DC susceptibility accompanied by a divergence between the ZFC and FC curves [19]. However, there is a striking difference between spin-glass and spin-ice freezing with respect to the distribution of relaxation times, which can be modelled at a given temperature using Cole-Cole analysis [26]. The $\chi'$ and $\chi''$ data for peak **1** of the AC susceptibility measurement without extra DC field, as well as the data for peak **2** with 200 Oe of DC field, were fitted within the Cole-Cole formalism, given by equation [19]:

$$\chi''(\chi') = -\frac{\chi_T - \chi_S}{2\tan\left[(1-\alpha)\frac{\pi}{2}\right]} + \sqrt{(\chi' - \chi_S)(\chi_T - \chi') + \frac{(\chi_T - \chi_S)^2}{4\tan^2\left[(1-\alpha)\frac{\pi}{2}\right]}} \quad (1)$$

Here $\chi_T$ is the isothermal susceptibility and $\chi_S$ is the adiabatic susceptibility. The parameter $\alpha$ represents the width of the distribution of relaxation times, where for a single spin relaxation time $\alpha = 0$. The values of $\alpha$ obtained for peak **2** ($\alpha \approx 0.7$, 0 Oe DC field, Fig. S7) and peak **1** ($\alpha \approx 0.56$, 200 Oe DC, Fig. S8) lie in the expected range for glass-like behavior, where the relaxation time typically covers a broad range of several orders of magnitude [19,27]. In contrast, spin ices exhibit either a single spin relaxation or a narrow range of relaxation times with an extremely low value of $\alpha \approx 0.001$ [28].

In addition, a quantitative measure of the frequency dependence of the maximum in $\chi'(T)$ can be estimated by the Mydosh parameter $\delta$ [29]:

$$\delta = \frac{\Delta T_f}{T_f \times \Delta(\ln(\omega))} \quad (2)$$

Here $T_f$ is the freezing temperature, the frequency is $\omega = 2\pi f$, and $\Delta T_f$ is the difference between the maximum and minimum values of $T_f$. The value of $\delta$ distinguishes a spin-glass state [19,30-32] ($0.001 < \delta < 0.08$) from a non-interacting ideal superparamagnet with much larger $\delta$ values [33,34]. The values of the Mydosh parameter for $K_3CrO_4$ ($\delta = 0.006$, 0 Oe DC field; $\delta = 0.017$, 200-400 Oe DC field) correspond to the intermediate situation of a cluster glass (CG), also known as a reentrant spin glass, for which $\delta \sim 0.01$-$0.09$ is expected [35-38].

These results suggest that the maxima in $\chi_{AC}$ are associated with randomly arranged, interacting magnetic clusters which become frozen below certain temperatures. Their origin most probably relates to the relaxation of ferromagnetic or ferrimagnetic clusters, but not antiferromagnetic clusters because in all cases $\chi''(T) \neq 0$ [25]. Moreover, it seems that there is a paramagnetic to cluster glass transition at ~36 K from the AC measurement without extra DC field, but with application of a small additional DC field that transition is to a ferromagnetic or glassy canted antiferromagnetic state and is followed at ~20 K by another transition to a cluster glass state. For such spin glass-like freezing occurring below a ferromagnetic transition it has previously been shown that the DC magnetization first increases with lowering temperature due to the appearance of the ferromagnetic state, but then decreases at even lower temperatures due to freezing [39,40].

### 3.4. Magnetization versus applied field

The nature of the magnetic states in $K_3CrO_4$ was next probed by means of magnetization (M) versus applied field (H) curves. The M-H curves exhibit an "S" shape at all temperatures below 60 K (Fig. 2(c); Fig. S9), indicating an uncompensated magnetic moment. However, the magnetization does not reach saturation (expected 1 $\mu_B$ / $Cr^{5+}$ cation) up to the highest applied field of 7 T at any temperature, which excludes a fully ferromagnetic state. There is a hysteresis loop below 20 K (Fig. 2(c); Fig. S9). Glassy compounds can show these features: the absence of magnetic saturation and weak hysteresis in the frozen state due to competing ferromagnetic and antiferromagnetic exchange interactions [35,37,38]. Furthermore, the M-H curves at 3-4 K exhibit a double pinched hysteresis loop (a closer view is shown in the left-hand panel of Fig. 4(a)). This might be associated with the magnetic transition that gives a maximum at 3 K in both the ZFC and FC DC magnetization measurements (Fig. 2(a)).

Moreover, signatures of field-induced transitions are observed in the temperature range 3-10 K (Fig. 4(a)). These transitions are temperature-dependent. The small anomaly at ±3.5 kOe at 3 K (marked with red triangles in the d$M$/d$H$ plot of Fig. 4(a)) shifts to higher magnetic field with increasing temperature (±4.75 kOe at 4 K; ±6 kOe at 6 K). The main peak in the d$M$/d$H$ plot at ±4 kOe at 3 K (marked with blue triangles in Fig. 4(a)) both shifts to lower field and becomes narrower with increasing temperature (±3.25 at 4 K; ±2 kOe at 6 K; ±1 kOe at 10 K).

### 3.5. Decay of thermoremanent magnetization

To investigate the mechanism by which the system decays back to equilibrium after an external magnetic field is applied, time-dependent thermoremanent magnetization measurements were carried out at 3 and 7 K (Fig. 2(d)). At higher temperatures, no significant relaxation was observed; the magnetization decayed immediately to zero on measurement time-scales. The data collected at 3 and 7 K do not follow a power-law decay or a simple logarithmic dependence. However, the curves can be fitted well with a stretched exponential function:

$$M(t) = M_0 + M_r \, exp\left[-\frac{t^{1-n}}{\tau}\right] \quad (3)$$

This law has widely been used to describe the magnetic relaxation in different glassy systems [41,42]. Here $M_0$ is the maximum magnetization at the start of the measurement and other components are related to the observed relaxation effect, where $M_r$ is a glassy contribution; the time constant $\tau$ and parameter $n$ are associated with the relaxation rate. The parameter $n = 0$ corresponds to a single time-constant and there is no relaxation at $n = 1$. The fitted curves match the data well, with the following extracted parameters at 3 K: $M_0 = 49.61$ emu/mol, $M_r = 18.67$ emu/mol, $\tau = 1776.12$ s, $n = 0.33$. The corresponding parameters at 7 K are as follows: $M_0 = 15.04$ emu/mol, $M_r = 5.65$ emu/mol, $\tau = 1508.85$ s, $n = 0.57$.

### 3.6. Dynamic scaling

More detailed insight into the dynamics of the two temperature-induced magnetic transitions in $K_3CrO_4$ can be obtained by further analysis of the $\chi_{AC}$ measurements. The frequency dependence of $\chi'(T)$ can be described by the critical slowing down formula [43] from dynamic scaling theory:

$$\tau = \tau_0 \times \left[\frac{T_f - T_g}{T_g}\right]^{-z\nu} \qquad (4)$$

Here $T_f$ is taken as the position of peak **2** at ~ 20 K (for measurements with 200-400 Oe applied DC field) and from the position of peak **1** at ~ 36 K (for the measurement with zero DC field) in $\chi'(T)$ for a given frequency $f$. $T_g$ is the temperature at which the maximum in the ZFC DC susceptibility is observed (29.7 K for 0 Oe DC field; 20 K for 200 Oe and 18 K for 400 Oe), because $T_g$ can be regarded as the value of $T_f$ for infinitely slow cooling ($\lim_{f \to 0} T_f$) [35]. The characteristic relaxation time of the dynamic fluctuations $\tau$ corresponds to the observation time $t_{obs} = 1/\omega = 1/(2\pi f)$ with the attempt frequency $\omega$, and the shortest time $\tau_0$ corresponds to the microscopic flipping time of the fluctuating entities. According to dynamic scaling theory, $\tau$ is related to the spin correlation length, $\tau \propto \xi^z$, and $\xi$ diverges with temperature as $\xi \propto [T_f/(T_f - T_g)]^\nu$ with the dynamic exponent $z$ and the critical exponent $\nu$ [44].

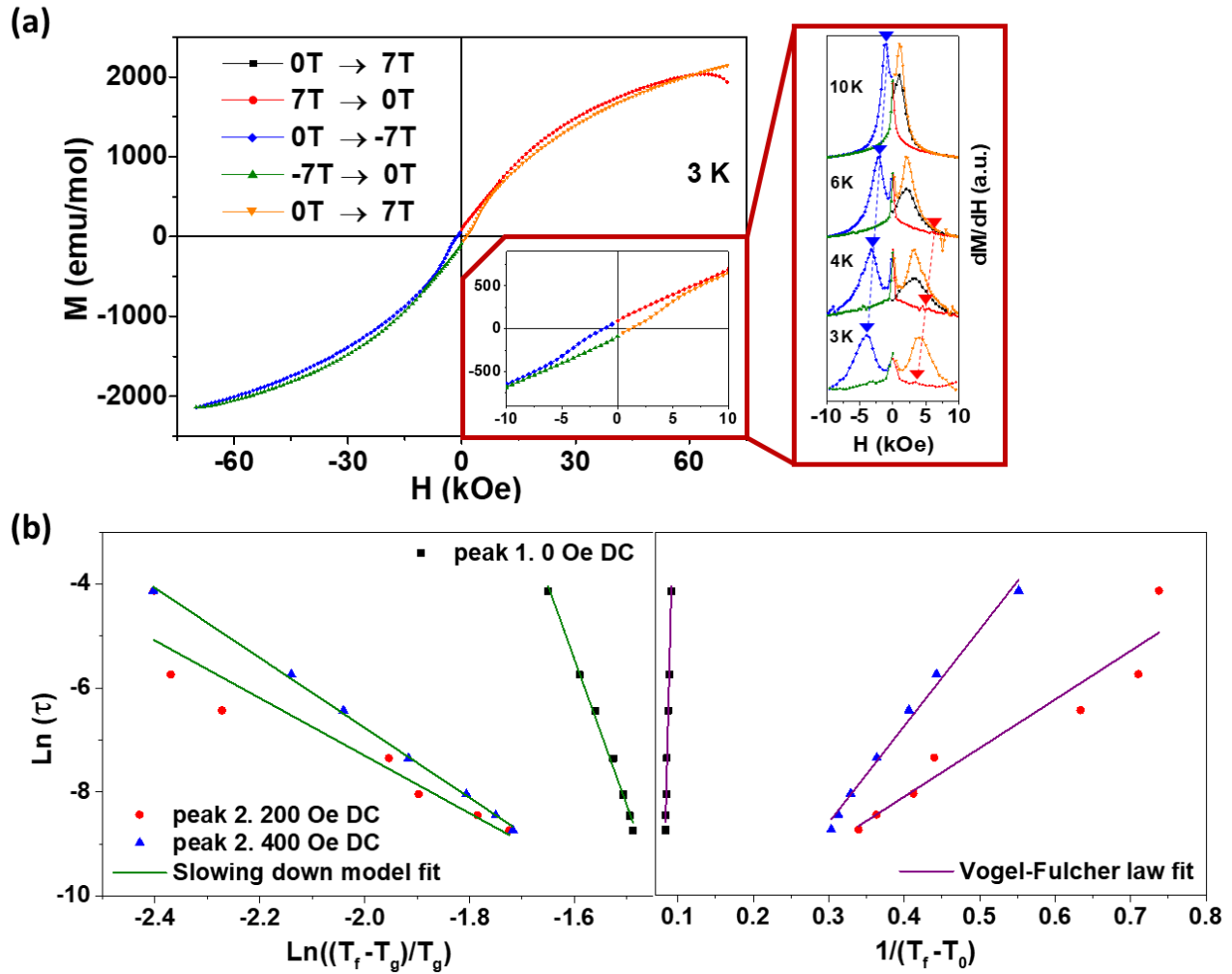

Fig. 4. (a) Magnetization vs applied DC field measurement at 3 K and field dependence of the susceptibility d$M$/d$H$ calculated from M-H curves at 3-10 K. The colors indicate the sequence in which data were collected (black, red, blue, green, orange). Triangles with dotted lines in the d$M$/d$H$ plots show the shifts of field-induced magnetic transitions at different temperatures. (b) Fits to maxima in $\chi'(T)$ for different frequencies for peak **2** (measurements in 200 / 400 Oe DC field) and peak **1** (measurement in zero DC field), using the slowing down formula (Eq. (4)) and the Vogel-Fulcher law (Eq. (5)) for $K_3CrO_4$.

The left part of Fig. 4(b) shows a linear fit of $ln\tau$ vs $ln((T_f-T_g)/T_g)$, allowing values of $zv$ and $\tau_0$ to be obtained. These parameters are given in Table I. The $zv$ values for peak **2** at 200-400 Oe DC field lie in the characteristic range for glassy magnetism $5 < zv < 13$ [45-47]. Typical values of $\tau_0$ for a canonical SG [47] lie in the range of $\sim 10^{-12}$–$10^{-14}$ s; for cluster glasses [35,41,48] with slower dynamics, $\tau_0$ is in the range $\sim 10^{-9}$–$10^{-11}$ s. The characteristic relaxation time of peak **2** in 200-400 Oe DC field is of the order of $\sim 10^{-9}$. These parameters strongly suggest that a magnetic cluster glass transition occurs at the temperature of peak **2** when an additional DC field is applied. Peak **1** at $\sim 36$ K in 0 Oe DC field has dynamics that are too fast even for a canonical SG and the $zv$ value is also too high.

The dynamic magnetic properties of a glassy system can also be described by the Vogel-Fulcher law [35,49], proposed for magnetically interacting clusters:

$$\tau = \tau^* \times exp\left[\frac{E_a}{k_B \times (T_f - T_0)}\right] \quad (5)$$

Here $T_0$ is a measure of the inter-cluster interaction strength, and $T_0$ is known as the Vogel-Fulcher temperature [29] and corresponds to the "ideal glass" temperature. Close to $T_0$, the Vogel-Fulcher law can be adjusted to match the power-law over a large frequency range [47]: $n\lfloor(40 \, k_B T_f)/E_a\rfloor \sim 25/zv$. This equation gives $E_a/k_B$ as listed in Table I. These values allow the data to be fitted using Eq. (5) (right-hand panels of Fig. 4(b)), yielding the parameters $\tau^*$, $T_0$ (Table I). The extracted $\tau^*$ values for peak **2** in 200-400 Oe DC field lie in the range of $\sim 10^{-7}$–$10^{-13}$ s anticipated for glassy bulk systems [29,35,50]. The obtained $\tau^*$ for peak **1** in zero DC field indicates that the dynamics are too fast, with a probably unphysical value, similar to the analysis using the critical slowing down formula.

Table I. Dynamic magnetic properties of $K_3CrO_4$.

|  | Slowing down formula | | Vogel-Fulcher law | | |
| --- | --- | --- | --- | --- | --- |
|  | $\tau_0$ (s) | $zv$ | $E_a/k_B$ (K) | $\tau^*$ (s) | $T_0$ (K) |
| Peak **2**, 200 Oe | $7.4\times10^{-9}$ | 5.7 | 11 | $5.6\times10^{-6}$ | 20.3 |
| Peak **2**, 400 Oe | $1.1\times10^{-9}$ | 6.9 | 22 | $4.37\times10^{-7}$ | 17.6 |
| Peak **1**, 0 Oe | $1.3\times10^{-22}$ | 28 | 593 | $5.12\times10^{-26}$ | 24.5 |

## 4. CONCLUSIONS

We have synthesized phase-pure $K_3CrO_4$ and for the first time investigated its magnetic properties. A Curie-Weiss fit to the DC magnetic susceptibility shows that antiferromagnetic interactions dominate, although the S-shaped M-H curves indicate a small uncompensated magnetic moment, likely from a canting of the spins. A bifurcation of the ZFC-FC magnetic susceptibility occurs below 38 K with a distinct peak in the ZFC branch (30 K) that suggests the presence of magnetic irreversibility. In addition, another magnetic transition is observed with a maximum at 3 K for the ZFC and FC branches.

AC magnetic susceptibility measurements indicate the presence of two magnetic transitions. Moreover, depending on whether an additional DC field is applied, the positions of the peaks associated with both magnetic transitions become frequency-dependent for both the real and the imaginary components. The peak at $\sim 36$ K seemingly corresponds to a paramagnetic - glassy canted antiferromagnetic transition, whereas the peak at $\sim 20$ K likely originates from the formation of a ferromagnetic/ferrimagnetic cluster glass state. Thermoremanent magnetization decay measurements confirm that a cluster glass state forms below the glass freezing temperature (20 K). Further evidence for a cluster glass state is provided by fits of the AC susceptibility data to both the standard critical slowing-down formula and the Vogel-Fulcher law.

In addition, indications of field-induced temperature-dependent transitions at lower temperatures (3-10 K) are found from the magnetic field dependence of the susceptibility calculated from the M-H curves. Further investigation of the nature of the magnetic transitions in $K_3CrO_4$ will require mapping of the magnetic structure as a function of temperature and applied field, for example from neutron diffraction measurements.

## AUTHOR CONTRIBUTIONS

Liliia D. Kulish: Conceptualization, Methodology, Validation, Formal analysis, Investigation, Data Curation, Writing - Original Draft, Visualization, Project administration. Graeme R. Blake: Conceptualization, Methodology, Validation, Resources, Data Curation, Writing - Review & Editing, Supervision, Project administration.

## ACKNOWLEDGMENTS

This work was supported by the European Union's Horizon 2020 research and innovation program under Marie Sklodowska-Curie Individual Fellowship, grant agreement no. 833550. We would like to thank Prof. Maxim Mostovoy and Joshua Levinsky for valuable discussion during the development of this research project as well as Ing. Jacob Baas for technical support.